\documentstyle[aps,preprint,epsf,citesort]{revtex}
\tightenlines
\begin{document}
\title
{Water-like anomalies for core-softened models of fluids:\\ One dimension}

\author{ M. Reza Sadr-Lahijany, Antonio Scala, Sergey~V.~Buldyrev,
and H. Eugene Stanley}

\address{Center for Polymer Studies and Department of Physics\\
        Boston University, Boston, Massachusetts 02215.}

\date{Submitted to Phys.~Rev.~E --- 19 July 1999}

\maketitle

\begin{abstract}

We use a one-dimensional (1d) core-softened potential to develop a
physical picture for some of the anomalies present in liquid water.  The
core-softened potential mimics the effect of hydrogen bonding. The
interest in the 1d system stems from the facts that closed-form results
are possible and that the qualitative behavior in 1d is reproduced in
the liquid phase for higher dimensions.  We discuss the relation between
the shape of the potential and the density anomaly, and we study the
entropy anomaly resulting from the density anomaly. We find that certain
forms of the two-step square well potential lead to the existence at
$T=0$ of a low-density phase favored at low pressures and of a
high-density phase favored at high pressures, and to the appearance of a
point $C'$ at a positive pressure, which is the analog of the $T=0$
``critical point'' in the $1d$ Ising model.  The existence of point $C'$
leads to anomalous behavior of the isothermal compressibility $K_T$ and
the isobaric specific heat $C_P$.

\end{abstract}


%\begin{multicols}{2}
\section{Introduction}
\label{secint}

Water, the most common liquid on earth, is uncommon in many of its
properties. For example, water at ambient pressure has an anomalous
density behavior, as for $T~<~4^\circ C$ its density decreases upon
cooling.  The subject of liquid anomalies is not limited to the density
anomaly.  Two other anomalies are the increase of isothermal
compressibility $K_T$ (density fluctuations) and of the isobaric
specific heat $C_P$ (entropy fluctuations) upon cooling.  All these
anomalies occur in liquid water~\cite{Debenedetti-book} and some occur
also in other liquids~\cite {Debenedetti-book,Mishima98,Yoshimura}.

In their pioneering work, Stell and Hemmer investigated potentials that
have a region of negative curvature in their repulsive core~\cite{Stell}
(henceforth referred to as core-softened potentials) in relation to the
possibility of a new critical point in addition to the normal liquid-gas
critical point.  They also pointed out that for the one dimensional
($1d$) model with a long range attractive tail, the isobaric thermal
expansion coefficient, $\alpha_P\equiv V^{-1}\left(\partial V/\partial
T\right)_P$ (where $V,T$ and $P$ are the volume, temperature and
pressure) can take an anomalous negative value. Debenedetti et al.,
using general thermodynamic arguments, confirmed that a ``softened
core'' can lead to $\alpha_P<0$~\cite{Debenedetti}. Stillinger and
collaborators found $\alpha_P<0$ for a 3d system of particles
interacting by purely repulsive interactions---a Gaussian potential
\cite{Stillinger}.

In this work, we investigate the relation of a core-softened potential
to the liquid anomalies mentioned above.  {\it Ab initio}
calculations~\cite{Liquid-Metals} and inversion of structure factor
data~\cite{Liquid-Metals,Head-Gordon} revealed that a core-softened
potential can be considered a realistic first order approximation for
the interaction of many materials with anomalous behavior
\cite{Head-Gordon,Young}, even in the case of network forming anomalous
liquids~\cite{Debenedetti-book}. A recent work has showed that the
anomalous behavior of a $1d$ model can be reproduced in two dimensions
($2d$) as well~\cite{ssbs,new2d}.

Here we will provide the details necessary to understand the preliminary
results announced in~\cite{ssbs}.  Specifically, we investigate
thoroughly an exactly solvable $1d$ model in order to develop an
intuitive picture of how the core-softened potential can lead to all
three anomalies and to relate the occurrence of these anomalies to the
shape and parameters of the potential.  We also discuss the possible
existence of a second critical point and its relation to the parameters
of the model.

The core-softened potential that we study is (Fig.~\ref{uform})
\begin{equation}
\label{uform}
 u(r)=\left\{ \begin{array}{ll} 
	\infty &0<r<a \\ -\lambda \epsilon &a<r<b\\
	 -\epsilon &b<r<c\\
	0 &r>c \end{array} \right. 
\end{equation}
where $r$ is the particle separation. The potential is composed of a
``hard core'' of diameter $a$ which has a repulsive shoulder (henceforth
referred to as ``softened-core'') of width $(b-a)$ and depth $\lambda
\epsilon$, and an attractive well of width $(c-b)$ and depth $\epsilon$.
Unless specified otherwise our numerical calculations are for the choice
of values $a=1$, $b=1.4$, $c=1.7$, $\epsilon=2$, and
$\lambda=0.5$~\cite{commentv}. A similar potential was introduced by
Ben-Naim to model water anomalies \cite{BenNaim1,Cho}.

A continuous potential of an analogous form models the interaction
between clusters of strongly bonded pentamers of water~\cite{Canpolat}.
This type of interaction qualitatively reproduces the density anomaly of
water. At sufficiently low pressures and temperatures, nearest neighbor
pairs are separated by a distance $r\approx b$ to minimize the energy.
As temperature increases the system explores a larger portion of the
configurational space in order to gain more entropy. This includes the
penetration of particles into the softened core, which can cause an
anomalous contraction upon heating.

The relation between the density anomaly and the shape of the potential
can also be understood using the thermodynamic relation
\begin{equation}
\label{dvdt}
\left(\partial V/\partial T\right)_P=\beta^2\left(P\langle(\delta V)^2\rangle_{PT}+ 
\langle\delta V\delta E\rangle_{PT}\right)
\end{equation}
where $\beta \equiv 1/k_BT$ with $k_B$ being the Boltzmann factor, $E$
is the energy and $\langle...\rangle_{PT}$ is the thermodynamic average in a
constant $P$, constant $T$ ensemble. For a system with density anomaly
the left hand side of Eq.~(\ref{dvdt}) is negative at temperatures below
the temperature of maximum density $T_{MD}$.  The first term on the
right-hand side, proportional to the square of the volume fluctuations,
is always positive.  Thus below $T_{MD}$ the second term must be
negative, which requires anti-correlation between the fluctuations in
$E$ and $V$. This kind of anti-correlation exists for the core-softened
model when the fluctuations occur between the states with $r\approx b$
and the states with $r\approx a$ which have a larger energy but a
smaller volume. We conclude that the core-softened potential is a
candidate for generating the density anomaly.

The paper is organized as follows: In Sec.~\ref{secsol} we present the
exact solution for the Gibbs potential. Sections III--VI discuss the
anomalies in the density and entropy, and their response functions
compressibility and specific heat. In Sec.~VII we discuss interesting
analogies with the Ising model. Finally, Sec.~VIII interprets the
anomalies in terms of two different local structures.

\section{Exact Solution For Gibbs Potential}
\label{secsol}

To derive the Gibbs potential for the model, we choose $c<2a$ to restrict
the interactions to nearest neighbors. The $1d$ model is then exactly
solvable using standard methods~\cite{Takahashi,BenNaim1,Cho}. The
partition function is (Appendix~\ref{appz})
\begin{equation}
\label{Z6}
Z(T,P)=\frac{1}{(\Delta V)[\Lambda(T)]^N(\beta P)^2} [\Omega(T,P)]^{N-1}
\end{equation}
where $N$ is the number of particles, $\Delta V$ is a discretization
factor which is needed for keeping $Z$ dimensionless and does not enter
the equation of state and $\Lambda(T)$ is the De Broglie wavelength. Here
\begin{equation}
\label{defkappa}
\Omega(T,P) \equiv \int_0^\infty dx~ e^{-\beta Px}e^{-\beta u(x)}
\end{equation}
where $u(x)$ is the interaction potential.

The Gibbs potential $G(T,P)\equiv -k_BT\ln Z(T,P)$ is then
\begin{equation}
\label{F}
G(T,P)=-k_BT\left[
	(N-1)\ln\Omega(T,P)-N\ln\Lambda(T)
	-2\ln P -\ln \Delta V
\right].
\end{equation}
In the thermodynamic limit $N\rightarrow \infty$, the terms $2\ln P$
and $\ln \Delta V$ are negligible compared to the other extensive (order
$N$) terms.

Equation~(\ref{Z6}) is valid for any $1d$ system in which each particle
interacts only with its nearest neighbors.  If the interaction potential
is given by Eq.~(\ref{uform}), then
\begin{equation}
\label{Psi}
\Omega(T,P)=\frac{1}{\beta P}\Psi(T,P)
\end{equation}
where (Appendix~\ref{appz})
\begin{equation}
\label{kdef}
\Psi(T,P)\equiv\left(\phi^\lambda\theta_a+(\phi-\phi^\lambda)\theta_b+
(1-\phi)\theta_c\right)
\end{equation}
and $\theta_x(P,T)\equiv e^{-\beta Px}$, 
$\phi(T)\equiv e^{\beta \epsilon}$.

\section{Density Anomaly}
\label{secdens}

We calculate the equation of state using the definition $V\equiv
{\partial G(T,P)}/{\partial P}$.  From Eq.~(\ref{F}) we find that, in
the thermodynamic limit $N \to \infty$, the average ``volume'' (length
in $1d$) per particle is
\begin{equation}
\label{v}
v \equiv V/N 
=-\frac{k_BT}{\Omega(T,P)}\frac{\partial \Omega(T,P)}{\partial P}.
\end{equation}
Using 
Eqs.~(\ref{Psi})~and~(\ref{v}) we find the equation of state

\begin{equation}
\label{eqeos}
v=\frac{k_BT}{P}-
\frac{k_BT}{\Psi(T,P)}
\frac{\partial \Psi(T,P)}{\partial P}
\end{equation}
where
\begin{equation}
\label{eqkp}
\frac{\partial \Psi(T,P)}{\partial P}=-\beta\left(\phi^\lambda a\theta_a+
(\phi-\phi^\lambda)b\theta_b+(1-\phi)c\theta_c\right).
\end{equation}

In the high temperature limit where $\phi\rightarrow 1$ and $\theta
\rightarrow 1$, the equation of state (\ref{eqeos}) tends to
\begin{equation}
\label{eqeostinfty}
v=\frac{k_BT}{P}+a , \qquad \qquad (T\rightarrow\infty)
\end{equation}
the equation of state for an ``ideal gas'' of rods, that is a system of
non-interacting rods of length $a$.

At $T=0$, $v$ as a function of $P$ has a discontinuity at
\begin{equation}
\label{eqpu1d}
P=P_{\mbox{\scriptsize up}}\equiv\frac{(1-\lambda)\epsilon}{b-a},
\end{equation}
which we call the upper transition pressure. 
At $T=0$,

\begin{equation}
\label{eqeos02}
v=\left\{ \begin{array}{ll} 
	b&\qquad \qquad 
P<P_{\mbox{\scriptsize up}}\\
	a&\qquad \qquad
P>P_{\mbox{\scriptsize up}}
	\end{array}\right.  
\end{equation}

We will return to this first-order transition in Sec.~\ref{secphas}.

Next we study the equation of state for fixed pressure.  For each value
of $P$, we find the value of $v(T)$ using Eq.~(\ref{eqeos})~
(Fig.~\ref{figlvst}).  The isobars separate into two different groups:

(i) For $P>P_{\mbox{\scriptsize up}}$, $v$ increases monotonically with
$T$, starting from its minimal value $v=a$. At higher pressures, the
nearest neighbors are pushed inside the softened core and, as a result,
the density anomaly does not occur.

(ii) When $P$ is lowered just below $P_{\mbox{\scriptsize up}}$, $v=b$
at $T=0$ [Eq.~(\ref{eqeos02})], and the $v(T)$ isobars show a maximum at
a temperature of minimum density $T_{mD}$ and a minimum at $T_{MD}$
\cite{BenNaim1,Cho,Yoshimura,Bell,Stillinger}.  For
$P<P_{\mbox{\scriptsize up}}$, the system starts at $T=0$ from the
bottom of the energy well at $v=b$.  Upon heating, the particles first
start to explore the wider region in the bottom of the potential well
and the system expands. At higher temperatures, the particles penetrate
the softened core and the system shrinks, showing an anomalous
temperature-driven contraction.  For even higher temperatures, the
particles move outside of the potential well, causing the system to
expand.

The minimum density point $T_{mD}$ is of interest. In real liquids with
$T_{MD}$, $T_{mD}$ is rarely observed~\cite{Yoshimura}, possibly because
it would be located at a very low temperature, where the liquid phase is
not stable.  As pressure is lowered further, the maximum and minimum
density points coincide at some point $(T_{\mbox{\scriptsize
low}},P_{\mbox{\scriptsize low}})$, which can be found from the system
of equations $\left( \partial v / \partial T \right)_P = 0 $ and $\left(
\partial^2 v / \partial T^2 \right)_P = 0 $.  We have observed this
behavior upon changing the parameters of the interaction potential (see.
e.g., Fig.~\ref{figlvst}b).  For $P<P_{\mbox{\scriptsize low}}$, no
density anomaly is observed, as shown in Fig.~\ref{figlvst}b.

\section{Isothermal Compressibility Anomaly}
\label{seccomp}
The isothermal compressibility is defined as
\begin{equation}
\label{defkt}
K_T\equiv-{1\over V}\left({\partial V\over\partial P}\right)_T={1\over
\rho}\left({\partial \rho\over\partial P}\right)_T.
\end{equation}
$K_T$ is thus the response of the volume to its conjugate variable
pressure and it is proportional to fluctuations in specific volume
\begin{equation}
\label{ktfluc}
K_T\;\propto\langle(\delta V)^2\rangle.
\end{equation}

For most materials $(\partial K_T/\partial T)_P>0$, so fluctuations
decrease upon cooling. In the case of water, for a wide range of
pressures, $K_T$ passes through a minimum and shows an anomalous
increase upon cooling. Along the $P=1$atm isobar, for example, the
minimum compressibility point is around $46^oC$~\cite{Debenedetti-book}.

From Eq.~(\ref{eqeos}) and Eq.~(\ref{defkt}), we find 
\begin{equation}
\label{eqkt}
K_T= \frac{1}{\beta v} 
\left( 
\frac{1}{P^2} + 
\frac{1}{\Psi(T,P)}
\frac{\partial^2\Psi(T,P)}{\partial P^2}
-\left(
\frac{1}{\Psi(T,P)}\frac{\partial\Psi(T,P)}{\partial P}
\right)^2
\right),
\end{equation}
where 

\begin{equation}
\label{eqkpp}
\frac{\partial^2 \Psi(T,P)}{\partial P^2}
=\beta^2\left(\phi^\lambda
a^2\theta_a+(\phi-\phi^\lambda)b^2\theta_b+(1-\phi)c^2\theta_c\right).
\end{equation}
Figure~\ref{figktvst1d} shows $K_T$ as a function of $T$ along isobars
from Eq.~(\ref{eqkt}).  As $T\rightarrow\infty$, $K_T$ tends to its
ideal gas value $1/P$, which is also predicted by
Eq.~(\ref{eqeostinfty}). As $T\rightarrow 0$, 

using Eq.~(\ref{defkt}), we find
\begin{equation}
\label{limktt0}
K_T\rightarrow 0  \;\;\;\;\; (T\rightarrow0, P\ne P_{\mbox{\scriptsize up}}).
\end{equation}
As $T$ decreases from infinity to zero, $K_T$ passes through two extrema,
a minimum and a maximum, between which $(\partial K_T/\partial T)_P<0$.
Thus the $1d$ core-softened model generates a compressibility anomaly.

\section{Density and compressibility extrema lines}
\label{ssectmd}

The locus of the points $T_{MD}$ in the $P-T$ phase diagram is of
special interest. For water, the shape of the $T_{MD}$ line helps
distinguish between different scenarios proposed to explain water's
anomalies \cite{Sciortino,Sastry,Speedy}.  In simulations
\cite{Pooletmd,Harrington,Harrington.spc3st2}, the $T_{MD}$ line
presents a ``nose,'' i.e., a point in which as $P$ decreases the slope
changes from negative to positive passing through an infinite value
(Fig.~\ref{figtmd1d}).

In Fig.~\ref{figtmd1d}, we present the $T_{MD}$ and $T_{mD}$ with the
same parameters as in Fig.~\ref{figlvst}.  We observe that both lines
originate at $P_{\mbox{\scriptsize up}}$~\cite{Meyer} and terminate at
$P_{\mbox{\scriptsize low}}$.  Moreover, the $T_{MD}$ line has a
negative slope for large $P$ which is in agreement with experimental
water and for lower $P$ the slope of the $T_{MD}$ line changes sign and
so a nose is present.

As shown in Fig.~\ref{figtmd1d}, the locus of extrema in compressibility
intersects the $T_{MD}$ line at its nose, as predicted by Sastry et
al.~\cite{Sastry} using thermodynamic relations. Moreover wherever the
$T_{MD}$ line has a negative slope, the compressibility must increase
upon cooling in the region to the left of the $T_{MD}$ line
\cite{Sastry}, resulting in a line of $K_T$ maxima which originates from
the point $C'(T=0, P=P_{\mbox{\scriptsize up}})$.

Noteworthy is the existence of a starting and ending point for the
$T_{MD}$ line. The starting point is the point $C'(T=0,
P=P_{\mbox{\scriptsize up}})$, and the ending point
$(T_{\mbox{\scriptsize low}},P_{\mbox{\scriptsize low}})$ where $T_{mD}$
and $T_{MD}$ meet.  In Fig.~\ref{figtmd1d}b we show that the overall
phase diagram does not change qualitatively upon varying the parameters
of the potential.

\section{Entropy Anomaly}

Since $(\partial S/\partial P)_T=-(\partial V/\partial T)_P$, if
$(\partial V/\partial T)_P<0$, then
\begin{equation}
\label{e20x}
\left({\partial S\over\partial P}\right)_T>0.
\end{equation}
Equation ({\ref{e20x}) is anomalous because, contrary to intuition,
compressing the system at constant $T$ increases its
entropy~\cite{FootNote1}. 
To calculate the entropy from Eq.~(\ref{F}), we use $S\equiv -(\partial
G/\partial T)_P$ 

We obtain
\begin{equation}
\label{S}
s\equiv\frac{S}{Nk_B}=\frac{3}{2}+\ln(\Psi(T,P)/(\Lambda\beta P))
-\frac{\beta}{\Psi(T,P)} 
\frac{\partial  \Psi(T,P)}{\partial \beta} 
\end{equation}
where 
\begin{equation}
\label{kdot}
\frac{\partial \Psi(T,P)}{\partial\beta}=
(\epsilon\lambda-Pa)\phi^\lambda\theta_a
+\epsilon(\phi-\lambda\phi^\lambda)\theta_b
-(\phi-\phi^\lambda)Pb\theta_b-\epsilon\phi\theta_c-(1-\phi)Pc\theta_c
.
\end{equation}

Using Eq.~(\ref{S}), we plot the entropy for two isotherms
(Fig.~\ref{Sisotherm}). For $T=0.6$, there is no density anomaly
(Fig.~\ref{Sisotherm}a) and hence no entropy anomaly. For $T=0.5$, there
is a density anomaly and hence an entropy anomaly; Fig.~\ref{Sisotherm}b
shows this anomalous increase of $S$ as a function of $P$.

\section{Analogies with the Ising model}
\label{sseckt}
\subsection{First Critical Point}

The gas-liquid first order transition line ending at a critical point
which is present in higher dimensions shrinks to the point
$C\equiv(T=0,P=0)$ for a $1d$ fluid in which the particles interact with
an attractive potential. This point $C$ is the remnant of what is a
critical point in higher dimensions: for example $K_T$ diverges for $T
\rightarrow 0$,
\begin{equation}
\label{e22a}
K_T\sim 1/T \qquad (T\to 0,P=0),
\end{equation}
analogous to the divergence of the magnetic susceptibility $\chi_T$
along the zero field line $H=0$ for the $1d$ Ising model, 
\begin{equation}
\label{e22b}
\chi_T\sim 1/T \qquad (T\to 0,H=0).
\end{equation}
The constant-pressure specific heat $C_P\equiv T(\partial S/\partial
T)_P$ is obtained from Eq.~(\ref{S}), with the result
\begin{equation}
\label{cp}
\frac{C_P}{Nk_B} =
\frac{3}{2} + \beta^2
\left(
	\frac{1}{\Psi(T,P)}
	\frac{\partial^2 \Psi(T,P)}{\partial \beta^2}
	-\left(
		\frac{1}{\Psi(T,P)}
		\frac{\partial \Psi(T,P)}{\partial \beta}
	\right)^2
\right).
\end{equation}
For small $P$ and $T$, the specific heat has the form 
\begin{equation}
\label{e24x}
C_P\sim a_0 + a_2(\beta P)^2
\end{equation}
the analog of the Ising case for small $H$ and $T$ \cite{chaikin-book} 
\begin{equation}
\label{e25x}
C_H \sim a_0 + a_2(\beta H)^2.
\end{equation}
These features are common to model 1d fluids with attractive potentials.

\subsection{Second Critical Point}

Next we will discuss the ``remnant'' of a second critical point for our
core-softened model.  In addition to the divergence along the $P=0$
isobar, we find a divergence along the $P=P_{\mbox{\scriptsize up}}$
isobar (Fig.~3),
\begin{equation}
\label{limktt0x}
K_T\sim \frac{1}{T} \qquad\qquad\qquad [T\rightarrow0,P=P_{\mbox{\scriptsize up}}].
\end{equation}

We also observe from Eq.~(\ref{eqeos02}) and Fig.~\ref{figlvst} that
there is a discontinuity in the order parameter $v$ when crossing 
$C'(P=P_{\mbox{\scriptsize up}},T=0)$
along the $T=0$ axis; this is the analog of what happens to the
magnetization when crossing the $H=0$ point along the $T=0$ axis for the
Ising model.

Next we consider $C_P$. 
Taking the limit $T\rightarrow 0$ of Eq.~(\ref{cp})
and defining $H\equiv |P-P_{\mbox{\scriptsize up}}|$, we find
\begin{equation}
\label{cpt0p}
C_P=\frac{3}{2}+\frac{wA^2}{(1+w)^2}(\beta H)^2,
\end{equation}
where we have introduced the parameters $A\equiv b-a$ and
$w\equiv\exp(-\beta A H)$. Eq.~(\ref{cpt0p}) is the analog of $C_H$ for
the Ising model.  Figure~\ref{figcp} shows the anomalous behavior of
$C_P$ as a function of $T$ along different isobars.

It is interesting to note that simulations of water using the ST2
potential display a compressibility anomaly due to the presence of a
second critical point in the metastable region of the liquid
\cite{pses92,Harrington}, and a line of $K_T$ maxima originates from
this second critical point.

\section{Interpretation in Terms of Two Local Structures}
\label{secphas}

The density and compressibility anomalies can be related to the
interplay between two local structures: an open structure in which the
nearest neighbor particles are typically at a distance $b$, and a
denser structure in which the nearest neighbors penetrate into the
softened core and are typically at a smaller distance $a$. The favored
local structure is determined by the Gibbs potential per particle 
\begin{equation}
\label{e1x}
g(T,P)=\mathop{\mbox{min}}_v\{u+Pv-Ts\},
\end{equation}
which is shown in Fig.~\ref{figfreeen} as a function of the volume per
particle, at $T=0$ for two different values of $P$. For
$P<P_{\mbox{\scriptsize up}}$ and $T=0$, the minimum corresponds the
``open structure'' with $r\approx b$.  

Increasing $P$ increases the value of $u+Pv-Ts$ for the open
structure (Fig.~\ref{figfreeen}). For $P>P_{\mbox{\scriptsize up}}$,
the minimum corresponds to the ``dense structure'' with $r\approx a$.

The value of $P_{\mbox{\scriptsize up}}$ of Eq.~(\ref{eqpu1d}) can also
be found by equating $u+Pv-Ts$ (at $T=0$) for the two local minima,
which results in
\begin{equation}
\label{eqpu}
P_{\mbox{\scriptsize up}}=-(u_{\mbox{\scriptsize
open}}-u_{\mbox{\scriptsize dense}})/(v_{\mbox{\scriptsize
open}}-v_{\mbox{\scriptsize dense}}).
\end{equation}
Substituting $u_{\mbox{\scriptsize open}}=-\epsilon,u_{\mbox{\scriptsize
dense}}=-\lambda\epsilon,v_{\mbox{\scriptsize
open}}=b,v_{\mbox{\scriptsize dense}}=a$ results in the same expression
as Eq.~(\ref{eqpu1d}).

For higher dimensions, Eq.~(\ref{eqpu}) helps to estimate the pressure
region in which a transition between a dense and an open structure could 
happen.

For $d=1$, the contribution of the $Ts$ term makes the double
well structure of Fig.~4 disappear when $T>0$. This may not be true in
higher dimensions. If we assume that the qualitative shape for $d>1$
changes little from the $d=1$ case, then for $d>1$ there can exist a
first-order transition line for small $T$, eventually terminating in a
critical point $C'$.

\section{Summary}

We used a $1d$ core-softened potential, which mimics the effect of
hydrogen bonding, to develop a physical picture for some of the
anomalies present in liquid water. We discussed the relation between the
shape of the potential and the anomalies in density and entropy and
their associated response functions $K_T$ and $C_P$. The form of the
potential leads to the existence at $T=0$ of a low-density phase
(favored at low pressures) and of a high-density phase (favored at high
pressures), and to the appearance of a point $C'$ at a positive
pressure, which is the analog of the $T=0$ ``critical point'' in the
$1d$ Ising model.

\section*{Acknowledgments}
\label{secack}

We thank M.~Canpolat, E.~La~Nave, M.~Meyer, S.~Sastry, F.~Sciortino,
A.~Skibinsky R.~J. Speedy, F.~W.~Starr, G.~S. Stell and D.~Wolf, for
enlighting discussions, and NSF for support.

\appendix

\section{Derivation Of the Free Energy for 1d Core-Softened Model}
\label{appz}

We start from the general definition of the partition function
\begin{equation}
\label{defZ}
Z(T,P)\equiv\frac{1}{(\Delta L)N!h^N}
\int_0^\infty dVe^{-\beta PV}
\int d^Npd^Nx ~e^{-\beta {\cal H} (p_1,...,p_N,x_1,...,x_N)}
\end{equation}
where $N$ is the number of particles, $V$ is the $1d$ system size, $x_i$
and $p_i$ are the position and momentum of particle $i$ respectively,
${\cal H}(p_1,...,p_N,x_1,...,x_N)$ is the Hamiltonian of the system.
$\Delta V$ is a discretization factor for $V$, which is needed for
keeping $Z$ dimensionless and does not enter any equation of state.  The
Hamiltonian can be partitioned into its kinetic and potential parts.
\begin{equation}
\label{defH}
{\cal H}
(p_1,...,p_N,x_1,...,x_N)\equiv \sum_{i=1}^N\frac{p_i^2}{2m}+U(x_1,...,x_N)
\end{equation}
One can separate Eq.~(\ref{defZ}) into a momentum and a configurational
integrals, where the momentum part is
\begin{equation}
\label{defZp}
Z_p(T)\equiv\frac{1}{h^N}\int d^Np\exp\left(-\beta \sum_{i=1}^N\frac{p_i^2}{2m}\right).
\end{equation}
This integral can be written as the product of $N$ integrals over
momenta and then using Gaussian integral formula ~\cite{formula-book} we
find
\begin{equation}
\label{Zp}
Z_p(T)=
\frac{1}{h^N}
\left( \int_{-\infty}^\infty dp~
\exp\left(-\beta p^2/2m\right)~ \right)^N
=\left[\Lambda(T)\right]^{-N}
\end{equation}
where $\Lambda(T)$, which has the dimension of length, is the
temperature dependent De Broglie wavelength $\Lambda(T)\equiv
h/\sqrt{2\pi mk_BT}$.  Thus the partition function takes the form
\begin{equation}
\label{Z1}
Z(T,P)=\frac{1}{(\Delta V)N!\Lambda(T)^N}\int_0^\infty
dVe^{-\beta PV}\int d^Nx e^{-\beta U(x_1,...,x_N)}.
\end{equation}

In order to take further steps, we use the fact that the range of
interaction is less that twice the hard core ($c<2a$). As a result, one
can think of the particles in the $1d$ system as a chain, in which each
particle interacts only with its nearest neighbors. Thus for each
arrangement of the particles, we can write the interaction potential $U$
as the sum of $N-1$ terms
\begin{equation}
\label{U}
U(x_1,...,x_N)=U(x_N-x_{N-1})+U(x_{N-1}-x_{N-2})+...+U(x_2-x_1)
\end{equation}
Using Eq.~(\ref{U}) we rewrite Eq.~(\ref{Z1}) as 
\begin{eqnarray}
\label{Z2}
Z(T,P)& =& \frac{1}{(\Delta V)\Lambda(T)^N}\int_0^\infty dVe^{-\beta
PV}\int_0^Vdx_N\int_0^{x_N}dx_{N-1}E(x_N-x_{N-1}) \nonumber \\
& & \int_0^{x_{N-1}}dx_{N-2}E(x_{N-1}-x_{N-2})
\cdots\int_0^{x_2}dx_1E(x_2-x_1),
\end{eqnarray}
where $E$ is defined as
\begin{equation}
\label{defE}
E(x)\equiv e^{-\beta U(x)}.
\end{equation}
In Eq.~(\ref{Z2}) we have used the symmetry of $Z$ under permutation of
the particles and thus have written Eq.~(\ref{Z2}) only for a specific
order of particles, in which $x_1<x_2<\cdots <x_N$. All other
permutations are equivalent to this integral, after renumbering the
particles. This results in a permutation factor $N!$ which has canceled
the $N!$ factor in the denominator.  We rewrite Eq.~(\ref{Z2}) as
\begin{equation}
\label{Z3}
Z(T,P)=\frac{1}{(\Delta V)\Lambda(T)^N}\int_0^\infty dVe^{-\beta PV}F_N(V)
\end{equation}
where we have introduced the function $F_i(x)$, defined recursively as
\begin{equation}
\label{FN}
F_N(V)\equiv\int_0^Vdx_NF_{N-1}(x_N)=(1*F_{N-1})(V)
\end{equation}
\begin{equation}
\label{Fi}
F_i(x_{i+1})\equiv\int_0^{x_{i+1}}dx_iE(x_{i+1}-x_i)F_{i-1}(x_i)=(E*F_{i-1})(x_{i+1})
\end{equation}
\begin{equation}
\label{F0}
F_0(x_1)\equiv 1.
\end{equation}
The star operator $*$, represents the binary convolution functional
defined as
\begin{equation}
\label{defconv}
(f*g)(x)\equiv\int_0^xdyf(x-y)g(y).
\end{equation}
We further use the notation ${\cal L}[f]$ for the Laplace transformation
functional which is defined as 
\begin{equation}
\label{defV1}
{\cal L}[f](z)\equiv\int_0^\infty e^{-zx}f(x)
\end{equation}
to find the following simple form for the partition function
\begin{equation}
\label{Z4}
Z(T,P)=\frac{1}{(\Delta V)\Lambda(T)^N}{\cal L}[F_N](\beta P).
\end{equation}
Note that according to Eqs.~(\ref{FN}),~(\ref{F0}) and ~(\ref{Z4})
\begin{equation}
\label{Z5}
Z(T,P)=\frac{1}{(\Delta V)\Lambda(T)^N}
{\cal L}[1*\overbrace{(E*(E*\cdots(E}^{N-1}*1)\cdots))](\beta P).
\end{equation}
Next we use the convolution theorem, which states that the Laplace
transform of the convolution is equal to the product of the Laplace
transforms of each function ~\cite{formula-book}
\begin{equation}
\label{lconv}
{\cal L}[f*g](z)={\cal L}[f](z)\times{\cal L}[g](z).
\end{equation}
and also the formula for the Laplace transform of a constant function 
~\cite{formula-book}
\begin{equation}
\label{l1}
{\cal L}[c](z)=\frac{c}{z}
\end{equation}
to obtain
\begin{equation}
\label{Z6p}
Z(T,P)=\frac{1}{(\Delta V)\Lambda(T)^N(\beta P)^2} [\Omega(T,P)]^{N-1}.
\end{equation}
Here $\Omega(T,P)$ is defined in Eq.~(\ref{defkappa}).

\section{Low-T Limit}
\label{ssecapp1}

In order to derive properties of the free energy at low temperature
approaching the point $C'(T=0,P=P_{\mbox{\scriptsize up}})$, we note
that in Eq.~(\ref{kdef}) as $T\rightarrow 0$, for small $P$, the
$\phi\theta_b$ term dominates, while for large $P$ the
$\phi^\lambda\theta_a$ dominates.  In order to find the limit of
$\Psi(T,P)$ and its derivatives, we note that
\begin{equation}
\label{0lims}
\frac{\phi^\lambda\theta_a}{\phi\theta_b}=\exp[{\beta(P(b-a)-(1-\lambda)\epsilon)}].
\end{equation}
The value of $P$ where the $\phi\theta_b$ term balances the
$\phi^\lambda\theta_a$ term follows an equating the argument of the
exponential to zero in Eq.~(\ref{0lims}). The result for
$P_{\mbox{\scriptsize up}}$ is given in Eq.~(\ref{eqpu1d}). Using
(\ref{eqpu1d}) and (\ref{0lims}), Eq.~(\ref{kdef}) becomes
\begin{equation}
\label{klimt0}
\Psi(T,P)\sim
\phi^\alpha\theta_x(1+\exp\left(-\beta(b-a)|P-P_{\mbox{\scriptsize
up}}|\right))
\end{equation}
where
\begin{equation}
\label{alfax}
\begin{array}{ll} 
	\alpha=1,\theta_x=\theta_b &\;\;(P<P_{\mbox{\scriptsize up}})\\
	\alpha=\lambda,\theta_x=\theta_a &\;\;(P>P_{\mbox{\scriptsize up}})\\
\end{array}
\end{equation}
For ${\partial \Psi(T,P)}/{\partial P}$ we use Eq.~(\ref{klimt0}) and
${\partial\theta_x}/{\partial P}=-\beta x\theta_x$ to find
\begin{equation}
\label{dpklimt0}
\frac{\partial \Psi(T,P)}{\partial P}\sim -(\beta x) \Psi(T,P),
\end{equation}
where $x=b$ for $b<P_{\mbox{\scriptsize up}}$ and $x=a$ for
$b>P_{\mbox{\scriptsize up}}$.  Using Eqs.~(\ref{alfax}),~(\ref{klimt0})
and~(\ref{dpklimt0}) in Eq.~(\ref{eqeos}) leads to Eq.~(\ref{eqeos02}).

To find the scaling form of $K_T$ for $T\to 0$, we use the equation for the
second derivative of $\Psi(T,P)$
\begin{equation}
\label{dppklimt0}
\frac{\partial^2 \Psi(T,P)}{\partial P^2}\propto (\beta x)^2 \Psi(T,P).
\end{equation}
>From Eq.~(\ref{eqkt}), we find $K_T$ for any $P\neq
P_{\mbox{\scriptsize up}}$ to behave as
\begin{equation}
\label{dppklimt0x}
K_T\sim (\beta v P^2)^{-1}=(P+\beta ~ x~P^2)^{-1} \;\;\;\;\;
(T\rightarrow0, P\ne P_{\mbox{\scriptsize up}})
\end{equation}
which is consistent with Eq.~(\ref{limktt0}).

In order to derive the limiting expressions for the entropy and specific
heat, we must differentiate the free energy with respect to $\beta$.  We
start by rewriting Eq.~(\ref{F}) as
\begin{equation}
\label{f}
g(T,P)\equiv\frac{G}{N}=k_BT\ln(\Lambda/\Omega)=k_BT\left(\frac{3}{2}\ln\beta
-\ln \Psi(T,P)+\ln P +constants\right).
\end{equation}
For $\Psi(T,P)$ we rewrite(~\ref{klimt0}) as 
\begin{equation}
\label{klimt0p}
\Psi(T,P)\sim E(T,P)\times(1+w(T,P)) \;\;\;\;T\rightarrow 0
\end{equation}
with the definitions
\begin{equation}
\label{Eedef}
\begin{array}{l} 
E(T,P)\equiv\exp\left(\beta (\epsilon\alpha-Px)\right)\\
w(T,P)\equiv\exp(-\beta)(b-a)|P-P_{\mbox{\scriptsize up}}|.
\end{array}
\end{equation}
Using these definitions we find
\begin{equation}
\label{dpklimt0x}
\begin{array}{l} 
{\partial \Psi(T,P)}/{\partial \beta}
\propto (\epsilon\alpha-Px)E\left(1+e(1-T)
\right)\\
{\partial^2 \Psi(T,P)}/{\partial \beta^2}
\propto (\epsilon\alpha-Px)^2E\left(1+e(1-T)^2\right)\\
\end{array}
\end{equation}
Using the above equations and Eq.~(\ref{cp}), we find
\begin{equation}
\label{cpt0}
C_P=\frac{3}{2}+\beta^2(\epsilon\alpha-Px)^2\left(\frac{1+w(1-t)^2}{1+w}-
\frac{(1+w(1-t))^2}{(1+w)^2}\right)
\end{equation}
whose leading term is Eq.~(\ref{cpt0p}).

%\end{multicols}

\begin{figure}[htb]
\caption{
General form for the core-softened potential studied here. The length
parameters $a,b,c$ and energy parameters $\epsilon,\lambda$ are shown.}
\label{figpotdisc}
\end{figure}

\begin{figure}[htb]
\caption{ 
Isobars of $v$, the average length per particle, for the discrete $1d$
core-softened potential showing the conditions under which a $T_{MD}$
and $T_{mD}$ exist. (a) The parameter values are
$a=1,b=1.4,c=1.7,\epsilon=2$ and $\lambda=0.5$. From
Eq.~(\protect\ref{eqpu1d}), these values result in $P_{\mbox{\scriptsize
up}}=2.5$. $P_{\mbox{\scriptsize low}}$ is almost zero, so no
$P<P_{\mbox{\scriptsize low}}$ isobar is shown. The $T_{MD}$ point is
marked by a filled ellipse and the $T_{md}$ point by an open
ellipse. (b) The parameter values are
$a=1,b=1.2,c=1.7,\epsilon=2,\lambda=0.5$ and thus from
Eq.~(\protect\ref{eqpu1d}), $P_{\mbox{\scriptsize up}}=5$. Now
$P_{\mbox{\scriptsize low}}\approx 3.3$, so a $P<P_{\mbox{\scriptsize
low}}$ isobar is shown.}
\label{figlvst}
\end{figure}

\begin{figure}[htb]
\caption{
Isothermal compressibility for the same parameter values as Fig.~2a. (a)
Isothermal compressibility along different isobars, with their maxima
marked by filled circles. $K_T$ along the isobar $P_{\mbox{\scriptsize
up}}$ diverges as $T\rightarrow 0$. (b) Log-log plot of the same results
showing the divergence of $K_T$ as $1/T$ along the critical isobar
$P=P_{\mbox{\scriptsize up}}$.}
\label{figktvst1d}
\end{figure}

\begin{figure}[htb]
\caption{
(a)The loci of the two density extrema ($T_{MD}$ and $T_{mD}$) and the
two $K_T$ extrema ($K_T^{\mbox{\scriptsize max}}$ and
$K_T^{\mbox{\scriptsize min}}$) for the discrete potential of Fig.~1
with the same parameter values as Fig.~2a, so $P_{\mbox{\scriptsize
up}}=2.5$ and $P_{\mbox{\scriptsize low}}\approx 0$. The locus of $K_T$
extrema has two ``loops,'' one connected to the point
$C'(T=0,P=P_{\mbox{\scriptsize up}})$ and the second to the point
$C(T=0,P=0)$. (b) Same for parameter values of Fig.~2b, for which
$P_{\mbox{\scriptsize up}}=5$ and $P_{\mbox{\scriptsize low}}\approx
3.3$. The two ``loops'' join.}
\label{figtmd1d}
\end{figure}

\begin{figure}[htb]
\caption{
(a) The region of the $P-T$ plane where density and entropy are
anomalous (grey); the parameters are the same as in Fig.~5a. (b) The
behavior of the entropy along two different isotherms. The $T=0.5$
isotherm intersects the anomalous region and shows the maximum and
minimum marked by the circles. The $T=0.6$ isotherm is outside the
anomalous region and does not show any anomaly.}
\label{Sisotherm}
\end{figure}

\begin{figure}[htb]
\caption{Constant-pressure specific heat along different isobars,
$P=2.7=P_{\mbox{\scriptsize up}}+0.2$ and $P=2.3=P_{\mbox{\scriptsize
up}}-0.2$. Near $T=0$, the two curves coincide, as predicted by
Eq.~(\protect\ref{cpt0p}), since the values of $H=|P-P_{\mbox{\scriptsize up}}|$
are identical.}
\label{figcp}
\end{figure}

\begin{figure}[htb]
\caption{
The function $u+Pv-Ts$ at $T=0$ for the core-softened potential of
Fig.~\protect\ref{figpotdisc}.  The equilibrium value of $v(P)$ is
determined as the absolute minimum, which is located at
$v=b$ for $P<P_{\mbox{\scriptsize up}}$ and at $v=a$ for
$P>P_{\mbox{\scriptsize up}}$. }
\label{figfreeen}
\end{figure}

\newpage

\begin{figure}[bt]
\narrowtext \centerline{
\hbox  {
        \vspace*{0.5cm}
        \epsfxsize=13.0cm
        \epsfbox{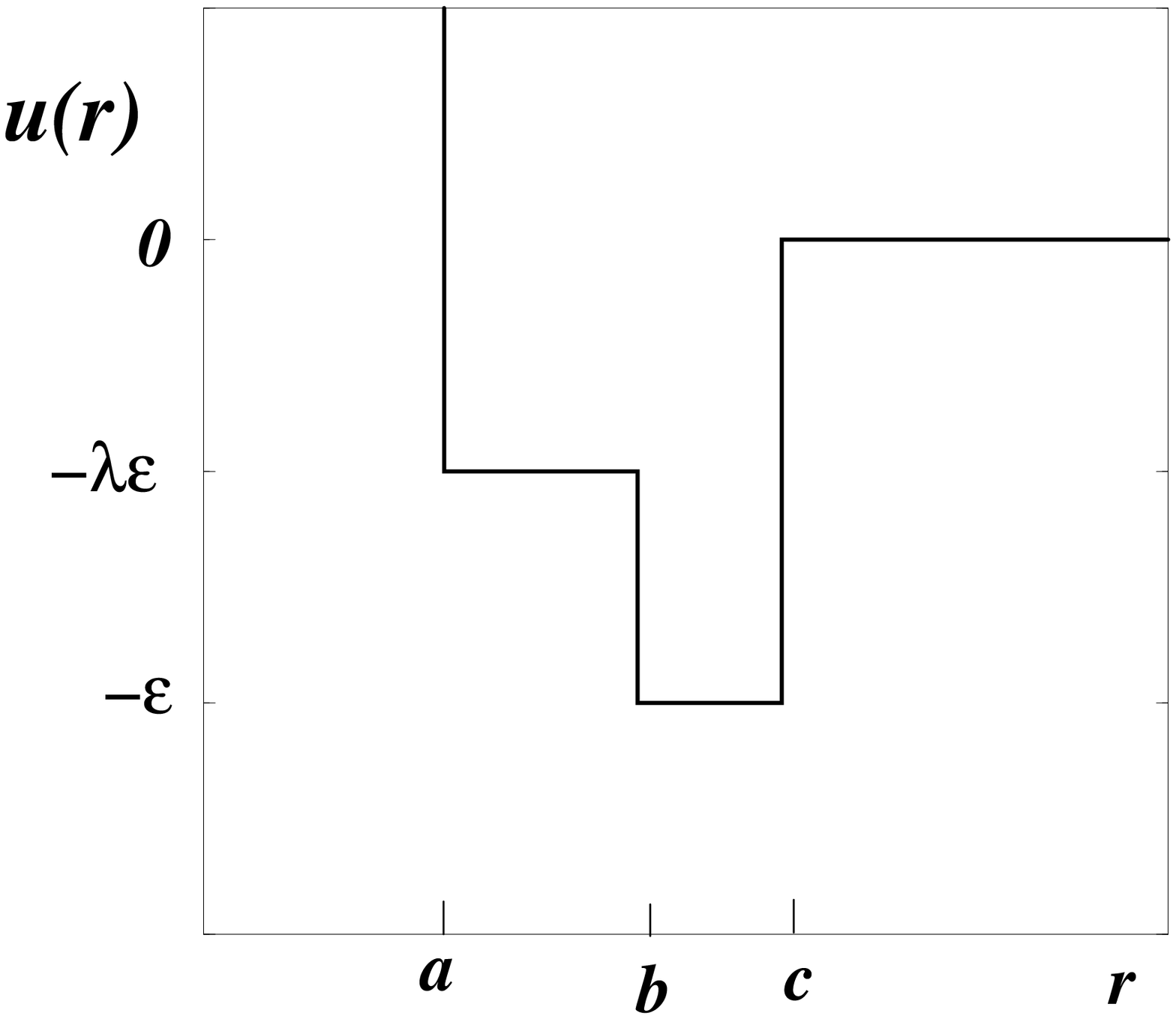}
        \vspace*{1cm}
       }
          }     
\end{figure}

\newpage

\begin{figure}[bt]
\narrowtext \centerline{
\hbox  {
        \epsfxsize=13.0cm
        \epsfbox{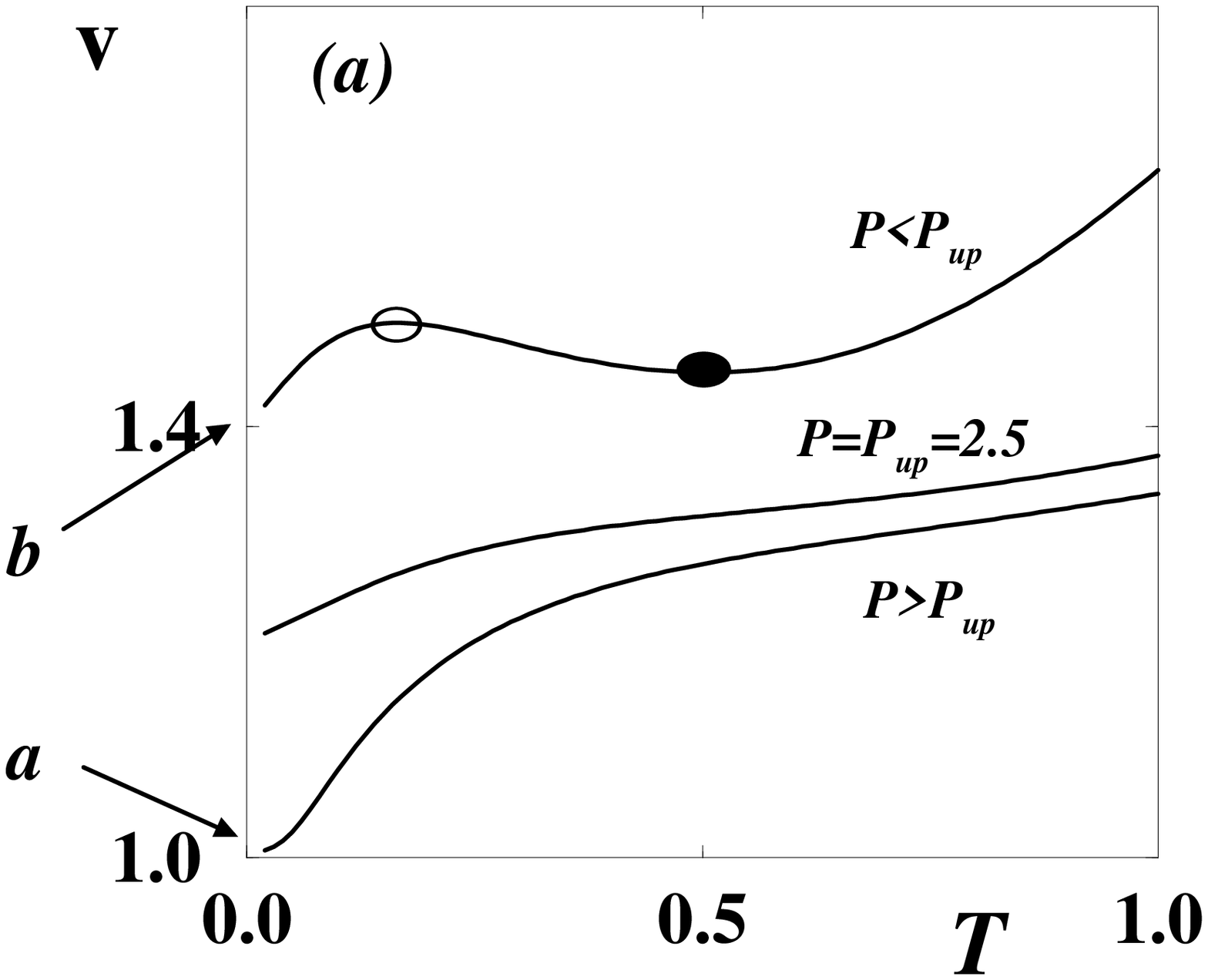}
	}
	}

%\newpage

\narrowtext \centerline{
\hbox  {
        \epsfxsize=13.0cm
        \epsfbox{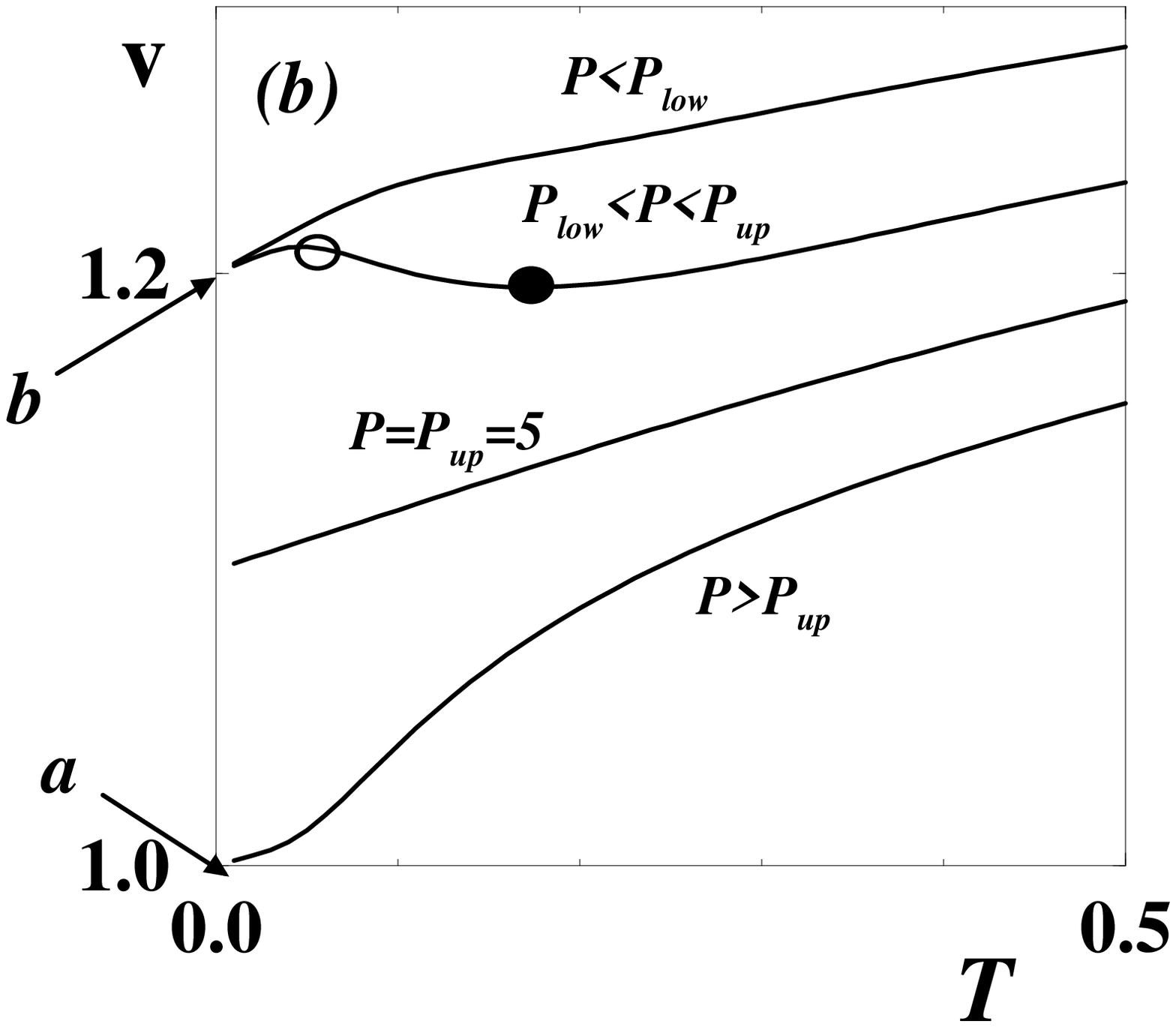}
        \vspace*{1cm}
	}
	  }
\end{figure}
\newpage

\begin{figure}[htb]
\narrowtext \centerline{
\hbox  {
        \epsfxsize=11.0cm
        \epsfbox{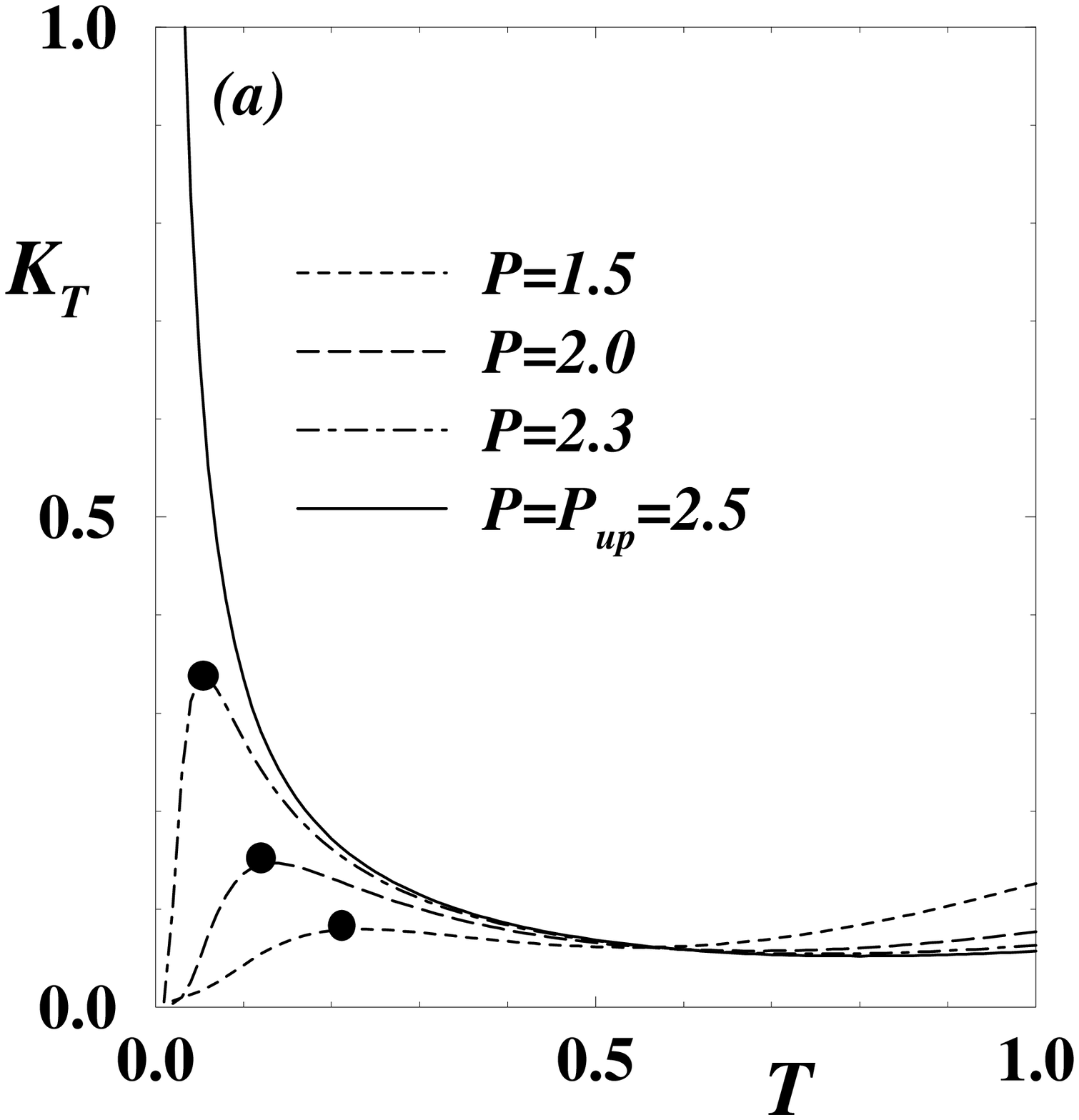}
       }
          }     

%\newpage

\narrowtext \centerline{
\hbox  {
        \epsfxsize=11.0cm
        \epsfbox{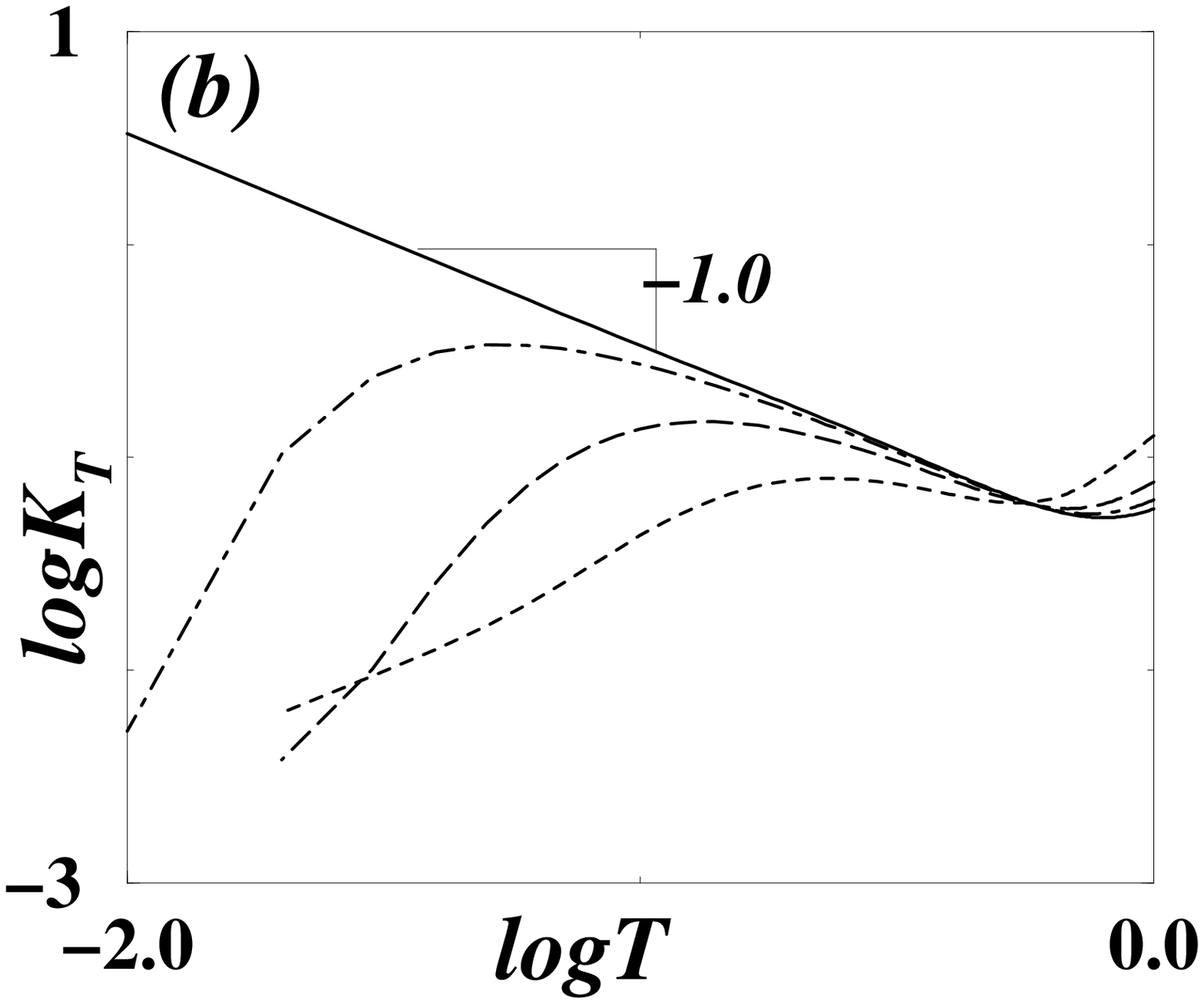}
        \vspace*{1cm}
       }
          }     
\end{figure}

\newpage

\begin{figure}[htb]
\narrowtext \centerline{
\hbox  {
        \epsfxsize=13.0cm
        \epsfbox{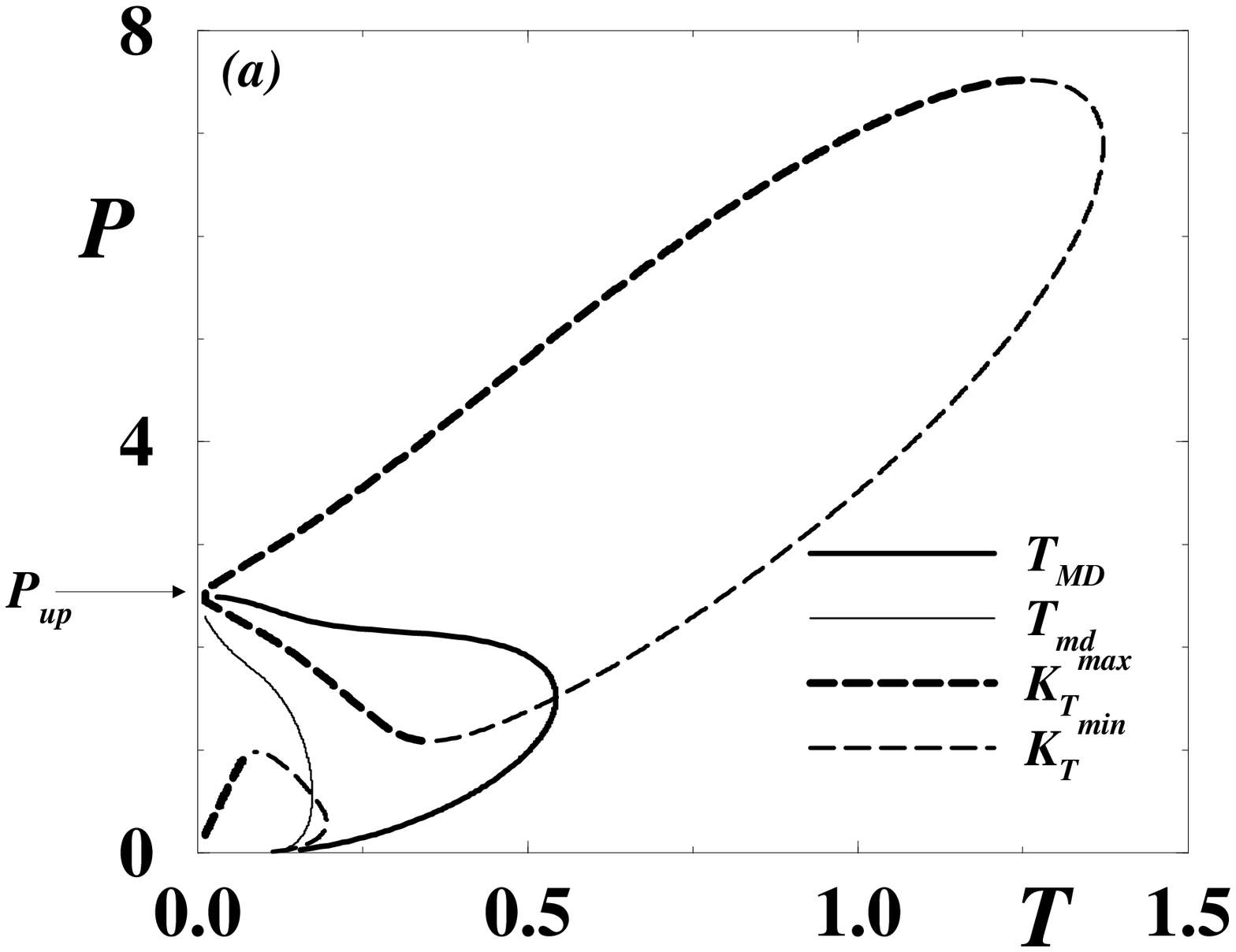}
       }
          }     

%\newpage

\narrowtext \centerline{
\hbox  {
        \epsfxsize=13.0cm
        \epsfbox{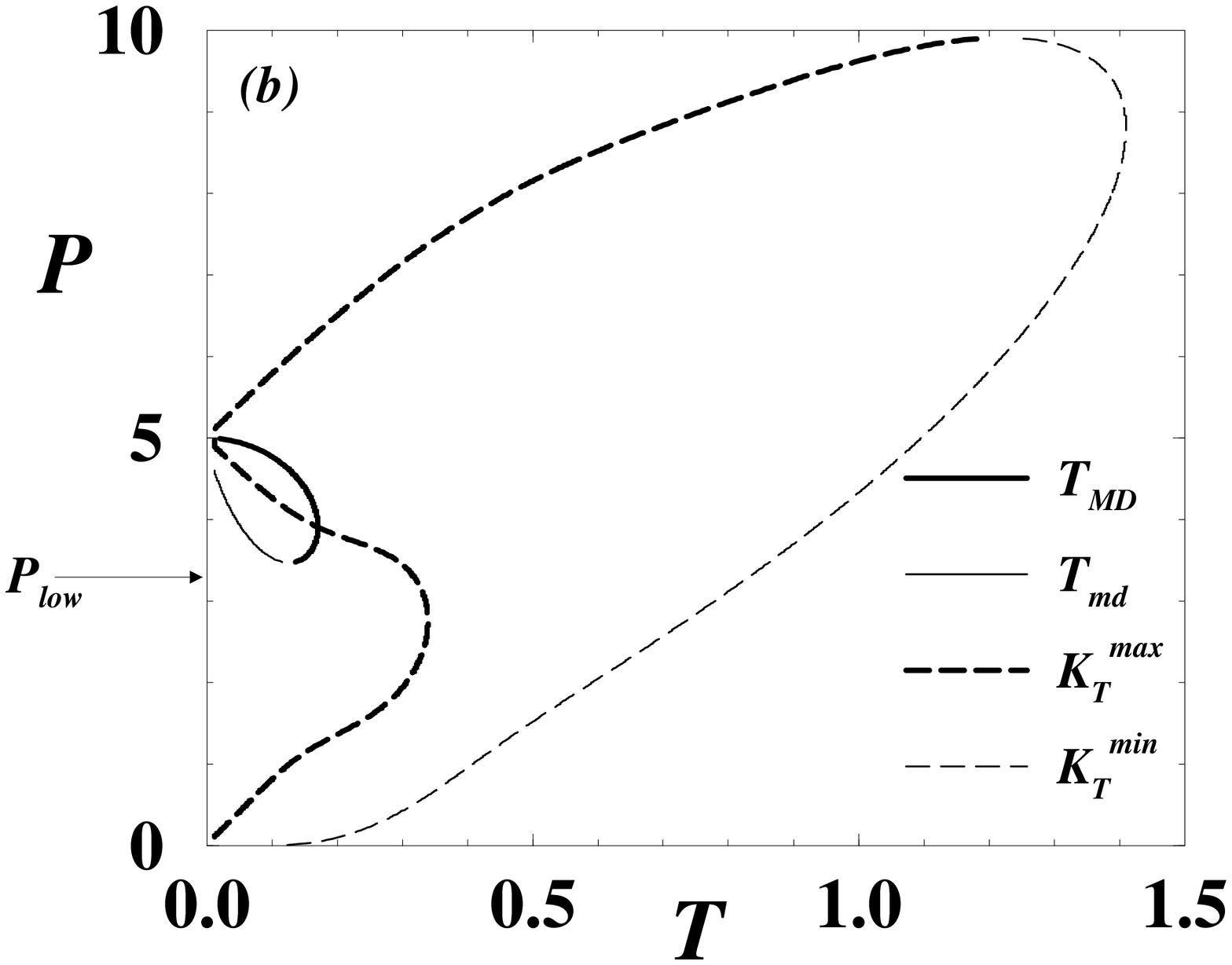}
        \vspace*{1.0cm}
      }
          }     
\end{figure}

\newpage

\begin{figure}[htb]
\narrowtext \centerline{
\hbox  {
        \vspace*{0.5cm}
        \epsfxsize=13.0cm
        \epsfbox{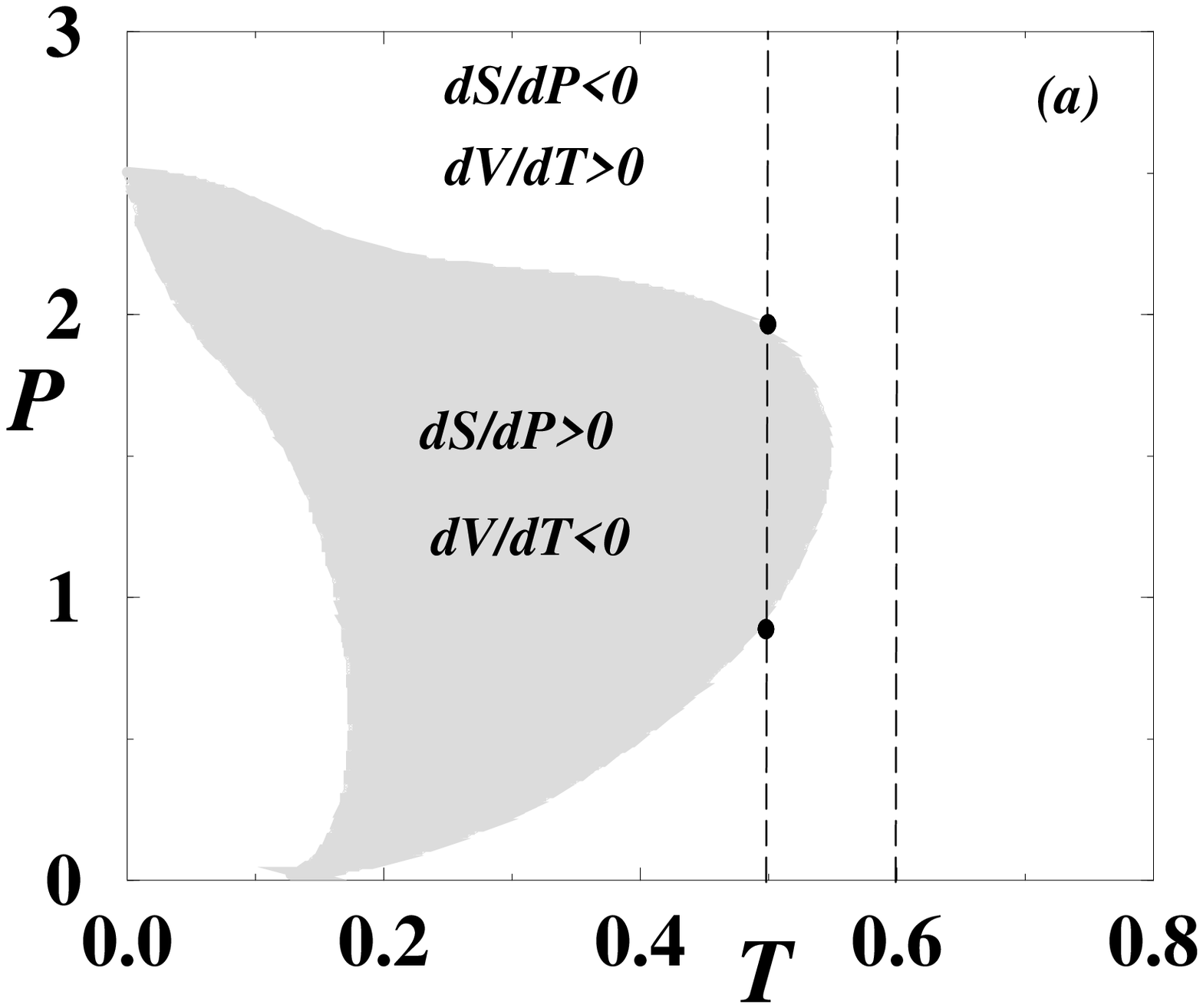}
        \vspace*{1.0cm}
       }
          }     
\end{figure}

\begin{figure}[htb]
\narrowtext \centerline{
\hbox  {
        \vspace*{0.5cm}
        \epsfxsize=13.0cm
        \epsfbox{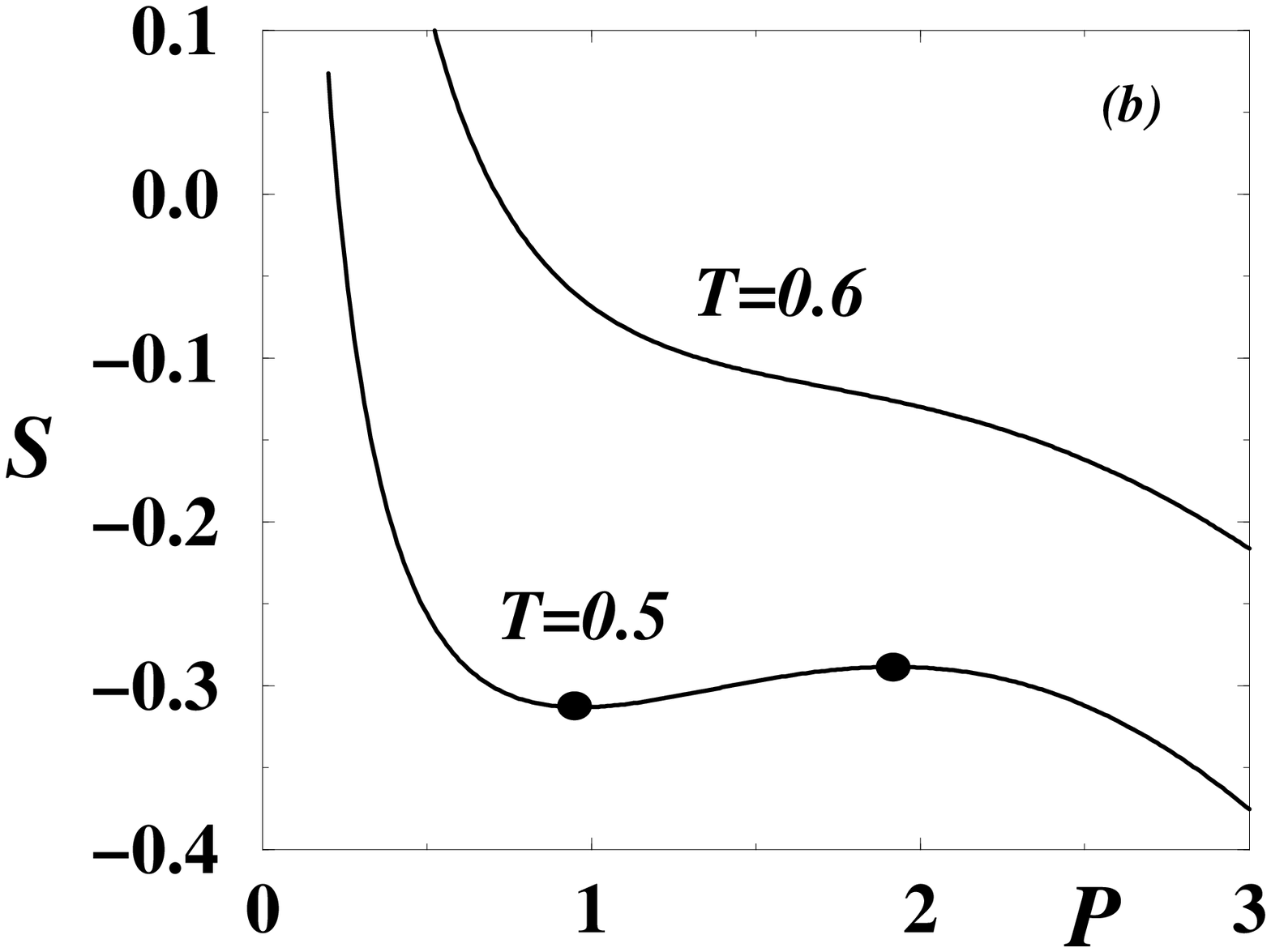}
        \vspace*{1.0cm}
       }
          }     
\end{figure}

\newpage

\begin{figure}[htb]
\narrowtext \centerline{
\hbox  {
        \vspace*{0.5cm}
        \epsfxsize=13.0cm
        \epsfbox{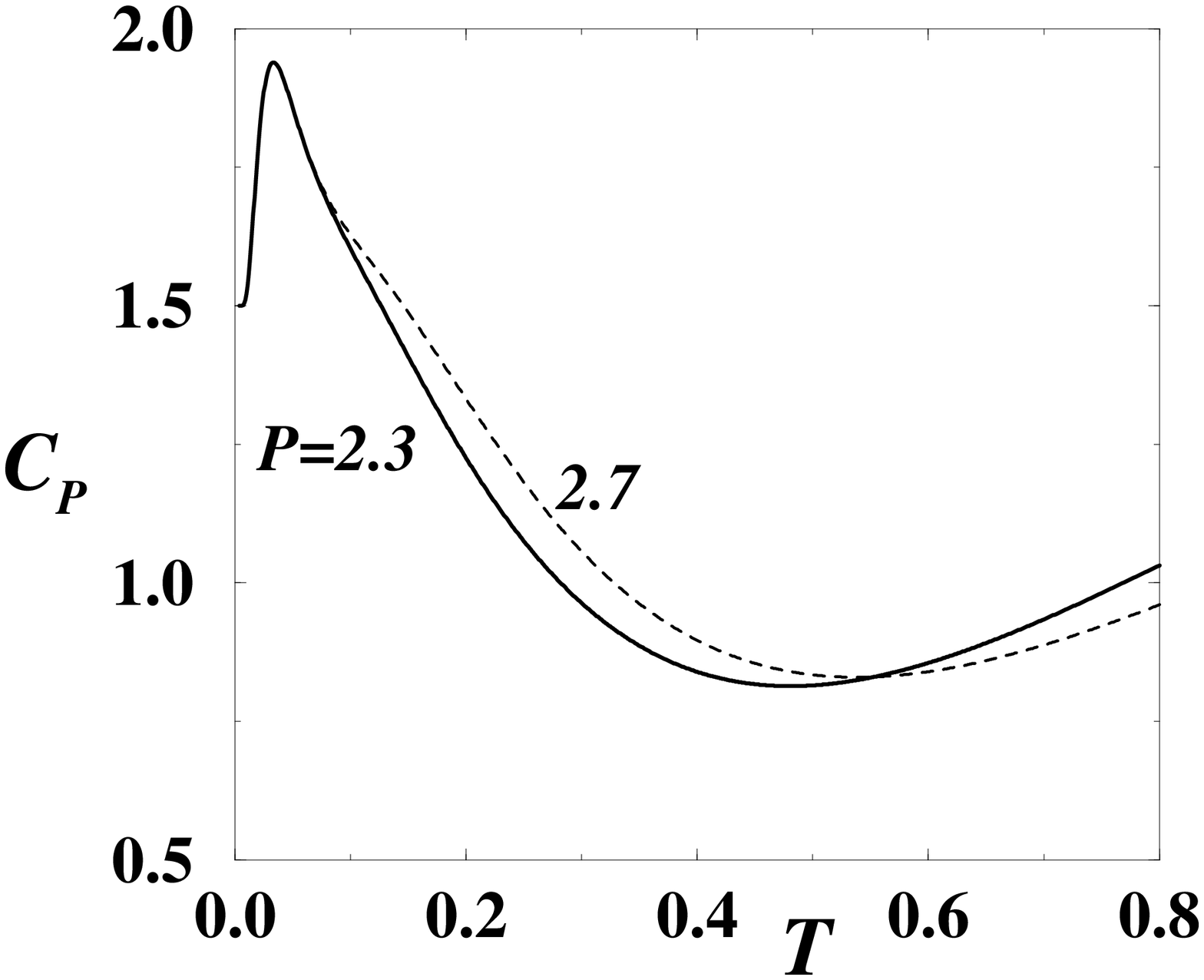}
        \vspace*{1.0cm}
       }
          }     
\end{figure}
\newpage

\begin{figure}[htb]
\narrowtext \centerline{
\hbox  {
        \epsfxsize=13.0cm
        \epsfbox{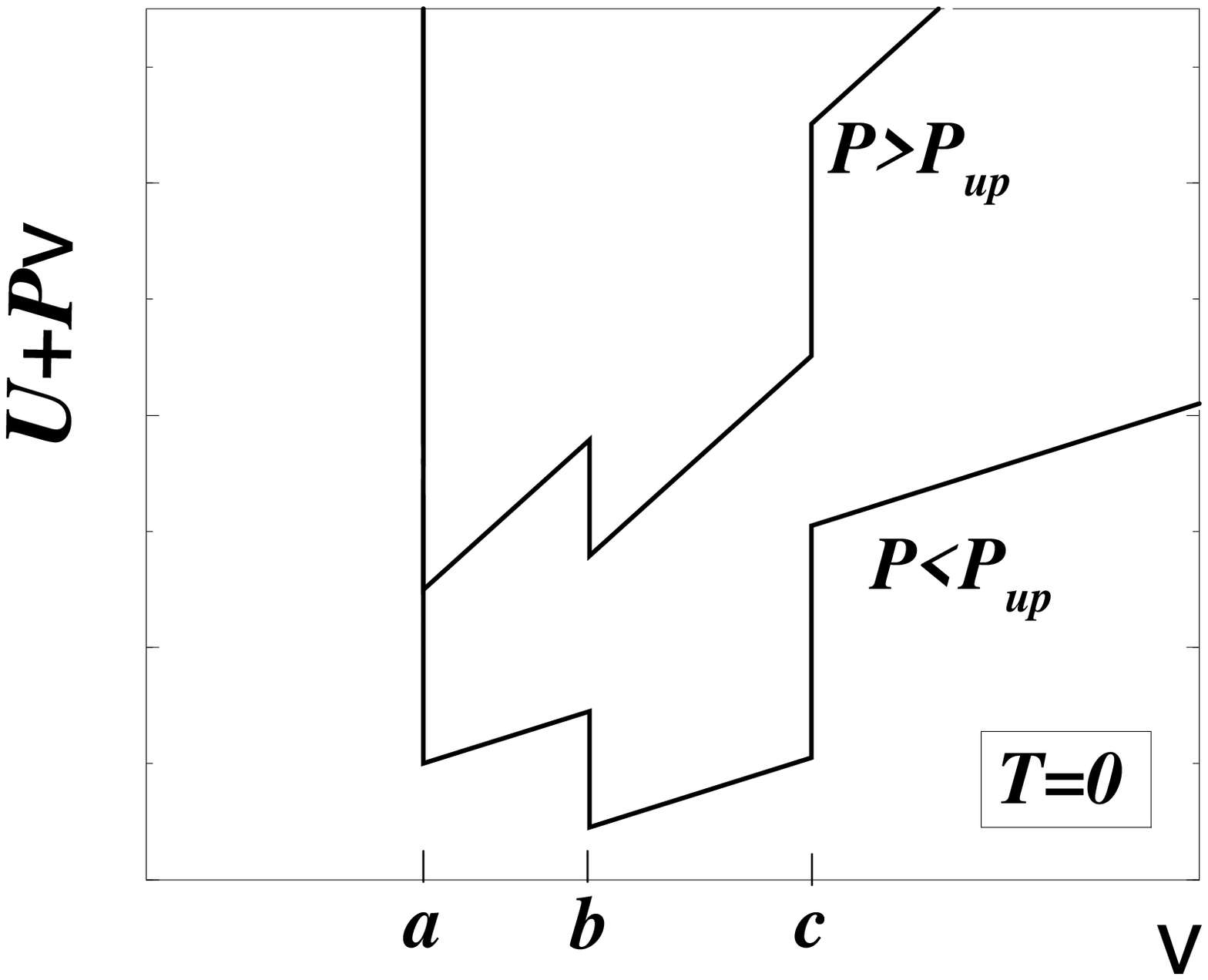}
       \hspace*{0.3cm}
        \vspace*{1cm}
       }
          }     
\end{figure}

%\end{multicols}

\begin{thebibliography}{99}
\bibitem{commentv} With these values, the discrete potential has a form
similar to the continuous version of the core-softened potential used in
$2d$~\protect\cite{ssbs}. Nevertheless the behavior we find holds for
any range of these parameters, as long as $0<a<b<c<2a$, $\epsilon >0$
and $\lambda <1$.
\bibitem{FootNote1}
For $d>1$, this can be accompanied by an increase of
the diffusion constant $D$. This anomalous increase in diffusion is
observed both in simulations \protect\cite{ssbs,Fstarr} and in water.
\end{thebibliography}
\end{document}